\documentclass[aps,twocolumn,showpacs,amssymb]{revtex4-1}
\date{\today}
\usepackage{graphicx}
\usepackage{color}
\begin{document}
\title{A monopole-optimized effective interaction for tin isotopes}
\author{Chong Qi}
\thanks{Email: chongq@kth.se}
\author{Z.X. Xu}
\address{Royal Institute of Technology (KTH), Alba Nova University Center,
SE-10691 Stockholm, Sweden}
\begin{abstract}
We present a systematic configuration-interaction shell model calculation on the structure of light tin isotopes with a new global optimized effective interaction. The starting point of the calculation is the realistic CD-Bonn nucleon-nucleon potential.  The unknown single-particle energies of the $1d_{3/2}$, $2s_{1/2}$ and $0h_{11/2}$ orbitals and the $T=1$ monopole interactions are determined by fitting to the binding energies of 157 low-lying yrast states in $^{102-132}$Sn. We apply the Hamiltonian to analyze the origin of the spin inversion between $^{101}$Sn and $^{103}$Sn that was observed recently and to explore the possible contribution from interaction terms beyond the normal pairing.
\end{abstract}
\pacs{21.30.Fe, 21.10.Dr, 21.60.Cs, 27.60.+j}

\maketitle
\section{Introduction}

Substantial experimental and theoretical efforts have been devoted in the past decade to explore the structure of light tin isotopes~\cite{Grawe01,Pin04,Banu05,Lid06,Kav07,Orce07,Vam08,Sew07,Doo08,Eks08,Dar10,Potel11,Jun11,East11,Mor11,All11,Pie11,Ast12,Gua12,Gua11,Gua08,Dik08,Wal11,Ata10,Loz08,Kum12}.  
The excitation energies of the first $2^+$ states in Sn isotopes between $^{102}$Sn and $^{130}$Sn are  established to possess an almost constant value (see, e.g., Refs. \cite{Grawe01,Talmi94,Rowe10}). This may be understood from the simple perspectives of generalized seniority scheme~\cite{Mor11,Talmi71,San97} and pairing correlation~\cite{And96,Dean03}.
A more realistic description of these nuclei requires a knowledge of the effective interaction between the valence nucleons that govern the dynamics~\cite{Talmi94,jen95}. The complete wave functions thus calculated show large overlap with those of the generalized seniority scheme, especially for the low-lying states of isotopes close to the $N=50$ and 82 shell closures~\cite{San97,Dean03}. The $J=0$ pairing channel of the effective interaction has been shown to play an essential role in reproducing the spacings between the ground states and $2^+_1$ states~\cite{Dean03}. Possible deviations from the generalized seniority scheme were suggested from $B(E2)$ measurements in the $2^+_1$ states \cite{Banu05}.

A microscopic shell-model description of the configurations of nuclei in the trans-tin region is a challenging task due to the scarceness of available experimental data and  the near-degeneracy in energy of the relevant $0g_{7/2}$ and $1d_{5/2}$ single-particle orbits~\cite{San95,Eng93,And96a}.  In the earlier shell-model calculations of Refs.~\cite{San95,Eng93}, the spacing between the two orbits were taken to be $\varepsilon(0g_{7/2})-\varepsilon(1d_{5/2}) = 0.2$~MeV and 0.5~MeV.
In Ref.~\cite{Fah01}, excited states in $^{103}$Sn have been observed using in-beam spectroscopic methods. The measured spectrum of $^{103}$Sn is very similar to those of $^{105,107,109}$Sn, with the spin-parity $5/2^+$ and $7/2^+$ for the ground state and first excited state. By a shell-model fitting procedure, the spacing between  $0g_{7/2}$ and $1d_{5/2}$ single-particle orbits was predicted to be $\varepsilon(0g_{7/2})-\varepsilon(1d_{5/2}) = 0.11$~MeV~\cite{Fah01}. It has long been expected that the ground state spin of $^{101}$Sn should be identical to those of $^{103-109}$Sn~\cite{Sew07} which can be approximately viewed as one-quasiparticle states~\cite{San95}.
However, in Ref.~\cite{Dar10}, the configurations of the ground state and first excited state in $^{101}${Sn} were determined to be $0g_{7/2}$ and $1d_{5/2}$, respectively. The spins of these states are reversed with respect to those in $^{103}${Sn}. 

Shell model calculations with empirical interactions have been very successful in explaining the structure and decay properties of light nuclei between $^{4}$He and $^{100}$Sn (see, e.g., Refs. \cite{ck,USD,Sch76,Poves81,hon02,hon09,Qi08a}) and heavier nuclei around shell closures \cite{Sch76,Cor09,Bro11}. The key to these calculations is a proper description of the monopole channel of the effective interaction \cite{ban64}, which determines the bulk properties of the effective interaction and governs the evolution of the effective single-particle energies (the mean field) as a function of valence neutron and proton numbers \cite{Poves81,Duf96}. The contribution of the monopole interaction becomes much more important with increasing valence nucleon numbers $N$ since it is proportional to $N(N-1)/2$. The light tin isotopes between shell closures $N=50$ and 82 are the longest chain that can be reached by contemporary shell model calculations. They may provide an ideal test ground to study the competition between different terms of the monopole interaction. 

Realistic effective interactions obtained from free nucleon-nucleon potentials provide a
microscopic foundation to shell model calculations \cite{jen95}. 
Extensive previous shell-model calculations tend to suggest that the realistic interaction can give a satisfactory description of the multipole part but not the monopole channel \cite{Poves81,USD,hon02,Yuan12}, which may be due to the lack of three-body forces \cite{Zuker03}. This is supported by recent shell-model calculations in Refs. \cite{Ots10,Holt12}, where it is shown that a better description of the oxygen and calcium chains can be obtained by including the three-body monopole interaction. Moreover, significant progress has been made in a variety of \textit{ab initio} calculations with three-body forces for light nuclei within the frameworks of Green’s function Monte Carlo \cite{Bri11}, no-core shell model \cite{Nav07,Roth11} and coupled-cluster \cite{Hag12} approaches.
For heavier nuclei, a more convenient approach is to treat the monopole interaction empirically \cite{Poves81,Duf96,zuk00}. Thus we are motivated to fine-tune the monopole part of the realistic interaction by fitting to available experimental data in tin isotopes. One may also get a limitation on the unknown single-particle energies of the orbitals $1d_{3/2}$, $2s_{1/2}$ and $0h_{11/2}$. We expect that the refined effective Hamiltonian will give a better understanding of the structure of trans-tin nuclei.

\section{Model space and optimization of the effective Hamiltonian}

We assume the doubly-magic $^{100}$Sn as the inert core. For the model space we choose the neutron and proton orbitals between the shell closures $N=Z=50$ and 82, comprising $0g_{7/2}$, $1d_{5/2}$, $1d_{3/2}$, $2s_{1/2}$ and $0h_{11/2}$. 
We also assume isospin symmetry in the effective Hamiltonian. A common practice in full configuration interaction shell model calculations is to express the effective Hamiltonians in terms of single-particle energies and two-body matrix elements numerically (see, e.g., the Oxbash Hamiltonian package \cite{Brown}),
\begin{eqnarray}
\nonumber H&=&\sum_{\alpha}\varepsilon_{\alpha}{\hat N}_{\alpha} \\
&&+ \frac{1}{4}\sum_{\alpha\beta\delta\gamma JT}\langle j_{\alpha}j_{\beta}|V|j_{\gamma}j_{\delta}\rangle_{JT}A^{\dag}_{JT;j_{\alpha}j_{\beta}}A_{JT;j_{\delta}j_{\gamma}},
\end{eqnarray}
where $\alpha=\{nljt\}$ denote the single-particle orbitals and $\varepsilon_{\alpha}$ stand for the corresponding single-particle energies. $\hat{N}_{\alpha}=\sum_{j_z,t_z}a_{\alpha,j_z,t_z}^{\dag}a_{\alpha,j_z,t_z}$ is the particle number operator. $\langle j_{\alpha}j_{\beta}|V|j_{\gamma}j_{\delta}\rangle_{JT}$ are the two-body matrix elements coupled to good spin $J$ and isospin $T$. $A_{JT}$ ($A_{JT}^{\dag}$) is the fermion pair annihilation (creation) operator. 
 For the model space we have chosen the effective Hamiltonian is such that it contains five single-particle energies and 327 two-body matrix elements. Among the two-body matrix elements there are 167 elements with isospin  $T=0$ and 160 elements with $T=1$. The number of matrix elements of a given set of $J$ and $T$ is given in Table \ref{ngp}. The monopole interaction is defined as the angular-momentum-weighted average value of the diagonal matrix elements $\langle j_{\alpha}j_{\beta}|V|j_{\alpha}j_{\beta}\rangle_{JT}$ for a given set of $j_{\alpha}$, $j_{\beta}$ and $T$ \cite{ban64,zuk00}. For the chosen model space there are 15 $T=0$ (and 1) monopole terms.

\begin{table*}
  \centering
  \caption{Numbers of two-body matrix elements $\langle j_{\alpha}j_{\beta}|V|j_{\delta}j_{\gamma}\rangle_{JT}$ for different sets of $J$ and $T$.}\label{ngp}
\begin{ruledtabular}
  \begin{tabular}{ccccccccccccc}
&$J=0$&$J=1$&$J=2$&$J=3$&$J=4$&$J=5$&$J=6$&$J=7$&$J=8$&$J=9$&$J=10$&$J=11$\\
\hline
$T=0$&$-$&36&16&48&16&25&11&9&3&2&$-$&1\\
$T=1$&15&6&46&18&34&13&16&6&4&1&1&$-$\\
  \end{tabular}
  \end{ruledtabular}
\end{table*}

The single-particle energies are assumed to be the same for all nuclei within the model space. They are given relative to the neutron $0g_{7/2}$ state. The energy of the $1d_{5/2}$ is taken as $\varepsilon (1d_{5/2})=0.172$ MeV \cite{Dar10}. The energies of other states have not been measured yet. They are adjusted to fit the experimental binding energies of tin isotopes. The single-particle energies of the proton orbitals are assumed to be the same as those of the neutron.

The starting point of our calculation is the realistic CD-Bonn nucleon-nucleon potential \cite{cdb}. The interaction was renormalized using the perturbative G-matrix approach, thus taking into account the core-polarization effects \cite{jen95}.  This interaction has been intensively applied in recent studies \cite{Banu05,Pro11,Back11}. The mass dependence of the 
effective interaction is not considered in the present work.

\begin{figure}[htdp]
\includegraphics[scale=0.35]{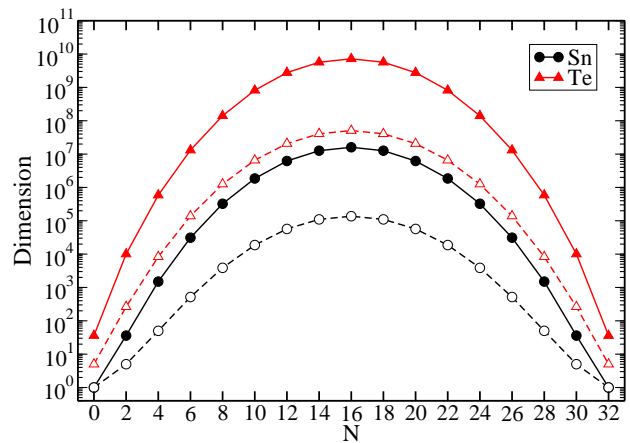}
\caption{(Color Online) Dimensions of the $M^{\pi}=0^+$ states in even-even Sn (circle) and Te (triangle) isotopes as a function of valence neutron numbers, where $M$ and $N$ denote the total magnetic quantum number and the number of valence neutrons, respectively. The open symbols connected by dashed lines give the dimensions of the corresponding $I^{\pi}=0^+$ states. \label{dimension}}
\end{figure}

Calculations are carried out within the so-called $M$-scheme where states with $M=I$ are considered. Diagonalizations are done with a parallel shell model program we developed a few years ago \cite{Qi08}.
Part of the calculations are checked with the shell model codes NuShellX \cite{rae} and ANTOINE \cite{ant} and that of the Oslo group \cite{Oslo}.  All the calculations are done on the computer Kappa at the National Supercomputer Center in Link\"oping, Sweden.  Matrices with dimensions up to $10^9$ (in the $M$-scheme) can be diagonalized with high efficiency. In Fig. \ref{dimension} we plotted the $M$-scheme dimensions for the $M=0$ positive-parity states in even-even Sn and Te isotopes. The dimensions of the corresponding $I^{\pi}=0^+$ states are also given for comparison. Only tin isotopes are considered in our fitting procedure due to the limitation in computing capability around mid-shell when both protons and neutrons are considered.

The experimental (negative) binding energy of a given state $i$ is given by
\begin{equation}
E^{\rm exp}_i = {\rm BE}^{\rm exp}_{\rm gs} (N) +{\rm Ex} (i),
\end{equation}
where BE and Ex denote the binding energy of the nucleus (with $N$ valence nucleons) and the corresponding excitation energy of the state relative to the ground state, respectively. The experimental data are taken from Refs. \cite{audi03,nudat}. A total number of $D=157$ yrast states from nuclei $^{102-132}$Sn are considered. 

We neglect isospin in following discussions for simplicity since the systems we handle in the present work only contain valence neutrons.
The calculated total energy of the state $i$ can be written as
\begin{equation}
E^{\rm cal}_i =C+N\varepsilon_{0}+\frac{N(N-1)}{2}V_m+ \langle\Psi_I|H|\Psi_I\rangle,
\end{equation}
where $\Psi_I$ is the calculated shell-model wave function of the state $i$ and $I$ is the total angular momentum.  The constant $C$ denotes the (negative) binding energy of the core $^{100}$Sn. The values of $\varepsilon_0$ and $V_m$ depend on the way the effective Hamiltonian is constructed. In the present work we assume $\varepsilon_{0g_{7/2}}=V_{m;0g_{7/2}^2}=0$ and the other single-particle energies and monopole interactions are given as relative values with respect to those of the orbital $0g_{7/2}$. Thus $\varepsilon_0$ and $V_m$ correspond to the real energy and monopole interaction of the $0g_{7/2}$ state in  $^{101}$Sn.

The excitation energy and wave function of a given state only depend on the shell model Hamiltonian $H$. One may rewrite the Hamiltonian as $H=H_m+H_M$ where $H_m$ and $H_M$ denote the (diagonal) monopole and Multipole Hamiltonians, respectively. The shell model energies can be written as
\begin{eqnarray}
\nonumber E^{\rm SM}&=&\langle\Psi_I|H|\Psi_I\rangle\\
\nonumber &=& \sum_{\alpha}\varepsilon_{\alpha}<\hat{N}_{\alpha}>+\sum_{\alpha\leq\beta}V_{m;{\alpha}\beta}\left<\frac{\hat{N}_{\alpha}(\hat{N}_{\beta}-\delta_{\alpha\beta})}{1+\delta_{\alpha\beta}}\right>\\
&&+\langle\Psi_I|H_M|\Psi_I\rangle,
\end{eqnarray}
where $\sum_{\alpha}<\hat{N}_{\alpha}> =N$ and
\begin{equation}
\sum_{\alpha\leq\beta}\left<\frac{\hat{N}_{\alpha}(\hat{N}_{\beta}-\delta_{\alpha\beta})}{1+\delta_{\alpha\beta}}\right> =\frac{N(N-1)}{2}.
\end{equation}

We optimize the single-particle energies and monopole terms of the realistic effective interaction by minimizing the quantity
\begin{equation}
\chi^2 = \sum_i^D\left[E_i^{\rm cal}-E_i^{\rm exp}\right]^2,
\end{equation}
where the summation runs over all states considered.  The quality of the fit can be measured by the root-mean square deviation as
\begin{eqnarray}
\sigma =\sqrt{\frac{\chi^2}{D-P}},
\end{eqnarray}
where $P$ denotes the total number of free terms that are to be considered in the fitting. As mentioned before, the single-particle energy of the orbital $1d_{5/2}$ are fixed at $\varepsilon (1d_{5/2})=0.172$ MeV \cite{Dar10}. One may rewrite the calculated total energy as
\begin{equation}
E^{\rm cal}_i =\sum_k^PV_kx_k + \varepsilon_{1d_{5/2}}<N_{1d_{5/2}}> + <H_M>,
\end{equation}
where $V_k$ denote the unknown single-particle energies and monopole interaction terms to be determined. The binding energy of the $^{100}$Sn core and the single-particle energy $\varepsilon_0$ are also taken as adjustable terms since the uncertainties in experimental data are still large \cite{audi03}. We have $P=20$ variables in total.

To minimize the $\chi^2$ function we apply a Monte Carlo global optimization method which we developed recently (denoted as MC in following discussions). It is an iterative approach. The basic idea is as follows: For the $n$-th fitting step we start with an initial set of $\{V_k(n)\}$ and
a new set of variables $\{V'_k(n)\}$ is proposed by the Monte Carlo sampling method. We require that $|V'_k(n)-V_k(n)|<\delta_k$ where $\delta_k$ denote the step lengths of the sampling. This new set will be accepted as $\{V_k(n+1)\}$ if we have $\chi^2(V'_k(n))<\chi^2(V_k(n))$.
Another consideration is that the proposed new set will also be accepted as $\{V_k(n+1)\}$ with certain probability $g$ even if one has $\chi^2(V'_k(n))>\chi^2(V_k(n))$.
This is a key part of the global optimization since the $\chi^2$ as a function of many variables may contain more than one minimum. Otherwise the global searching may be trapped in a local one. The step is repeated until convergence. The step lengths $\delta_k$ and the probability function $g$ can also be adjusted in the fitting to getting a faster convergence.

The advantage of the Monte Carlo global optimization method  is that no information on the derivatives (i.e., $x_k$ in the present study) is required. This is very convenient when other observables (e.g., $B(E2)$ values) are included in the fitting.

To get the mean-square deviations $\chi^2$ for a given set of $\{V_k(n)\}$ and the large number of succeeding samplings $\{V'_k(n)\}$ (of the order $10^2-10^3$) one has to re-diagonalize the corresponding shell-model Hamiltonian matrices.
Since the shell-model diagonalizations around $N=16$ are still time consuming, we further assume that the wave functions are stable against the variation of the effective Hamiltonian, $\langle\Psi_I(n)|\Psi'_I(n)\rangle\sim1$. One has
\begin{equation}
E^{\rm cal}_i(n)'-E^{\rm cal}_i(n) \approx\sum_k^P(V'_k(n)-V_k(n))x_k,
\end{equation}
from which the $\chi^2$ value for a given sampling can be calculated approximately in a straightforward way. The wave functions and coefficients $x_k$ are re-calculated when a new set of variables are accepted. This is known as the linear approximation based on which standard fitting approaches can be applied \cite{hon02,Bru77}. 

As a comparison, the singular value decomposition (SVD) approach is also employed in the fitting process. The SVD approach was used recently in Refs. \cite{Ber09,Joh10}.
For calculations with the SVD approach, the constants $C$ and $\varepsilon_0$ are taken as free parameters with no restriction. In the MC approach, we assume that $C$ and $\varepsilon_0$ can only take values within the range $-825.5\sim-8.24$ MeV and $-12\sim -8$ MeV, respectively, by considering the uncertainties in experimental data \cite{audi03}. For the monopole interactions we assume $|V_{m}|\leq1.5$ MeV. These restrictions can be adjusted in the fitting process if necessary.

The fitting is carried out in three steps. In the first step we only consider 131 states in the nuclei $^{102-112}$Sn and $^{120-132}$Sn. The nuclei $^{113-114}$Sn and $^{118-119}$Sn are considered in the second step to further fine-tune the effective interaction. The three isotopes $^{115-117}$Sn are added to the calculation in the last step. Our calculations show that convergence is already reached in the second step.

\begin{figure}[htdp]
\vspace{0.5cm}
\includegraphics[scale=0.35]{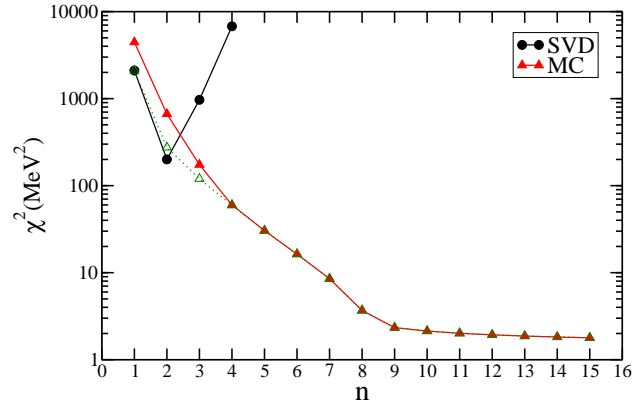}
\caption{(Color Online) The convergence of the mean-square deviations $\chi^2$ as a function of iteration number within the singular value decomposition (SVD) and Monte Carlo global optimization (MC) approaches starting from a random monopole Hamiltonian. See text for the difference between the red-solid and green-open triangles. \label{fit1}}
\end{figure}

To test the fitting approaches mentioned above, calculations are done with 10 sets of random monopole Hamiltonians. They are generated by the Monte Carlo sampling approach with the restriction 0.172 MeV $< \varepsilon_{\alpha}\leq$ 5 MeV and $|V_{m;\alpha\beta}|\leq 1$ MeV. Then these Hamiltonians are optimized by fitting to the 131 states mentioned above. We found that in all cases one can get convergence with the MC approach within ten iterations. As a typical example, one set of these calculations are plotted in Fig. \ref{fit1}. Two types of MC calculations are presented in the figure. The solid triangles correspond to the calculations with the restrictions on the constants $C$, $\varepsilon_0$ and $V_m$ mentioned above. These restrictions are removed in calculations marked by the open symbols. This is why in the first ($n=1$) iteration the $\chi^2$ value is smaller in the latter case. For $n=1$ calculations with the SVD approach give the same $\chi^2$ value as that of the second MC calculation. But the new set of variables $\{V_k(n=2)\}$ predicted by these approaches are very different.  Among calculations with the ten random samplings, as in Fig. \ref{fit1}, the SVD approach diverges in most cases. The reason for the divergence may be that the step lengths (the difference between $\{V_k(n+1)\}$ and $\{V_k(n)\}$) predicted by the SVD approach is too large that one can not apply the linear approximation. In the MC approach we restrict the step length be $\delta\leq1$ MeV.
This is a rational restriction by taking into account the fact the $T=1$ monopole interactions are mostely small and close to zero in shell-model calculations in light and medium-mass nuclei \cite{USD,hon09}.

Fig. \ref{fit1} suggests that our restrictions on the $C$, $\varepsilon_0$ and $V_m$ are reasonable and have no influence on the final results. 
It should also be mentioned that the monopole Hamiltonians that are fitted starting from these random samplings are very similar to each other.

\begin{figure}[htdp]
\includegraphics[scale=0.35]{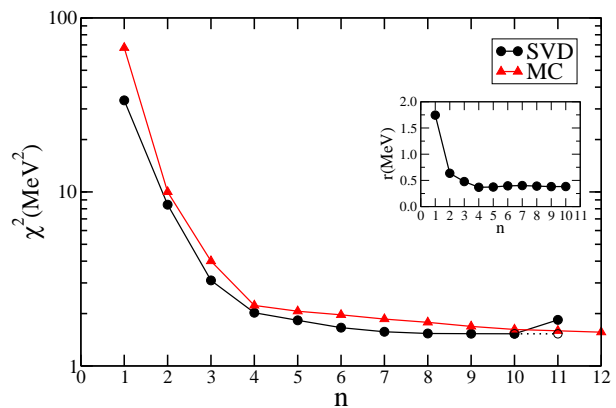}
\caption{(Color Online) The convergence of the mean-square deviations $\chi^2$ as a function of iteration number for fittings that are started from the realistic CD-Bonn interaction. The insert shows the maximal deviation $r={\rm max}|E_i^{\rm cal}-E_i^{\rm exp}|$ in each step.\label{fit2}}
\end{figure}

In Fig. \ref{fit2} the same calculations are done starting from the realistic CD-Bonn potential. The 131 yrast states are considered in the fitting. The initial root-mean-square deviation is around 750 keV. This is reduced to about 110 keV after 10 iterations. The $\chi^2$ given by the SVD approach is smaller than the MC method for the same reason as before. But the variables $C$, $\varepsilon_0$ and $V_m$ predicted by the SVD approach soon fall into the expected range after a few iterations. The $\chi^2$ value given by the SVD calculation starts to fluctuate around $n=10$. This is avoided in the MC calculations by gradually decreasing the maximal range $\delta$. In the calculation labeled by the open circle presented in the figure, the MC approach is applied instead of SVD.

In Ref. \cite{Ber05} the maximal deviation $r={\rm max}|E_i^{\rm cal}-E_i^{\rm exp}|$ is minimized in their fitting to experimental binding energies within the Skyrme-Hartree-Fock approach. In the insert of Fig. \ref{fit2} we plot the maximal deviation $r$. It also gradually decreases as a function of the iteration even though the criterion of our calculation is to find the minimum of $\chi^2$.

\section{Results and discussions}

The fitting approach described above has been successfully applied in deriving the effective interactions for several nuclear regions. In particular, we constructed a new interaction to describe the structure of the $N=83-85$ isotones by assuming the nucleus $^{146}$Gd as the core \cite{Had12}, which is not as good shell closure as $^{100}$Sn. As a result, the standard SVD approach failed to get a converging result.
The new interaction for light tin isotopes derived in the present work is briefly discussed below. It has already been used in recent studies on the level structure and electromagnetic transition properties of the odd-$A$ nucleus $^{109}$Te \cite{Pro12} and the E2 transition properties in Sn isotopes \cite{Back12}. 

\begin{table}
  \centering
  \caption{The final mean square deviation $\chi^2$ (in MeV$^2$) given by the SVD and MC fittings and the predicted values of the variables $C$, $\varepsilon_0$ and $V_m$ and their uncertainties (in MeV).}\label{table3}
\begin{ruledtabular}
  \begin{tabular}{ccccc}
&$\chi^2$&$C$&$\varepsilon_0$&$V_m$\\
\hline
MC&2.697&$-825.5 \pm  0.288$&$-10.671\pm 0.0410$&$0.167 \pm0.00249$\\
SVD& 2.686&$-825.5 \pm  0.288$&$-10.669  \pm 0.0410$& $0.172 \pm 0.00249$\\
  \end{tabular}
  \end{ruledtabular}
\end{table}

The monopole Hamiltonians we derived in Fig. \ref{fit2} are slightly refined by including the nuclei $^{113-119}$Sn into the fitting. As mentioned before, we include a total number of 157 states in the fitting among which there are 31 binding energies and 126 excitation energies. As seen in Table \ref{table3}, after around 15 iterations both calculations give a mean-square deviation $\chi^2\sim2.35$ MeV. It means that these states can be reproduced within an average deviation of about 123 keV. In Table \ref{table3} we also give the values of the variables $C$, $\varepsilon_0$ and $V_m$ predicted by the MC and SVD calculations. The uncertainties within these variables are also analyzed with the help of the SVD approach.
The  $C$ and $\varepsilon_0$ values predicted by the fitting are in reasonable agreement with the binding energies given in Ref. \cite{audi03}.

The largest uncertainties of the optimized monopole Hamiltonian are related to the single-particle energies. The values predicted by the MC approach are $\varepsilon_{ 1d_{3/2}}=5.013\pm3.10$, $\varepsilon_{2s_{1/2}}=0.369\pm2.63$ and  $\varepsilon_{ 0h_{11/2}}=3.249\pm0.83$ MeV.
The single-particle energies given by the SVD approach are  $\varepsilon_{ 1d_{3/2}}=5.182\pm3.16$, $\varepsilon_{2s_{1/2}}=0.409\pm2.73$ and  $\varepsilon_{ 0h_{11/2}}=3.236\pm0.84$ MeV. More experimental efforts are desired in order to get a better constraint on these single-particle energies.

In Table \ref{table2} we compare the optimized monopole terms from the SVD and MC approaches and those of the realistic CD-Bonn interaction. The list of the two-body matrix elements and part of the calculated results on tin isotopes with the optimized interaction can be found in Ref. \cite{sn100}.

\begin{table}
  \centering
  \caption{The strengths of the $T=1$ monopole interactions $V^m_{J;j_1j_2}$ (in MeV) of the original CD-Bonn potential and the interactions optimized by the SVD and Monte-Carlo fitting approach.}\label{table2}
\begin{ruledtabular}
  \begin{tabular}{ccccc}
$j_1$&$j_2$&CD-Bonn&SVD&MC\\
\hline
  $0g_{7/2}$ &  $0g_{7/2}$    &   0.000&   0.000&   0.000\\
 $1d_{5/2}$  &$1d_{5/2}$     &  -0.200&  -0.127&  -0.121\\
 $1d_{3/2}$  &$1d_{3/2}$     &  -0.105&   0.200&   0.179\\
 $2s_{1/2}$  &$2s_{1/2}$     &  -0.834&  -0.707&  -0.749\\
  $0h_{11/2}$ &  $0h_{11/2}$     &  -0.136&  -0.250&  -0.244\\
  $0g_{7/2}$ &$1d_{5/2}$     &  -0.129&  -0.157&  -0.151\\
  $0g_{7/2}$ &$1d_{3/2}$     &  -0.060&  -0.086&  -0.139\\
  $0g_{7/2}$ & $0h_{11/2}$     &  -0.251&  -0.287&  -0.261\\
  $0g_{7/2}$ &$2s_{1/2}$     &  -0.105&  -0.062&  -0.028\\
 $1d_{5/2}$  &$1d_{3/2}$     &  -0.231&  -0.726&  -0.607\\
 $1d_{5/2}$  &$2s_{1/2}$     &  -0.201&   0.228&   0.203\\
 $1d_{5/2}$  & $0h_{11/2}$     &  -0.106&   0.012&  -0.016\\
 $1d_{3/2}$  &$2s_{1/2}$     &  -0.134&  -0.783&  -0.768\\
 $1d_{3/2}$  & $0h_{11/2}$     &  -0.191&  -0.122&  -0.116\\
 $2s_{1/2}$   &$0h_{11/2}$     &  -0.141&  -0.018&  -0.013\\
  \end{tabular}
  \end{ruledtabular}
\end{table}

\begin{figure}[htdp]
\includegraphics[scale=0.35]{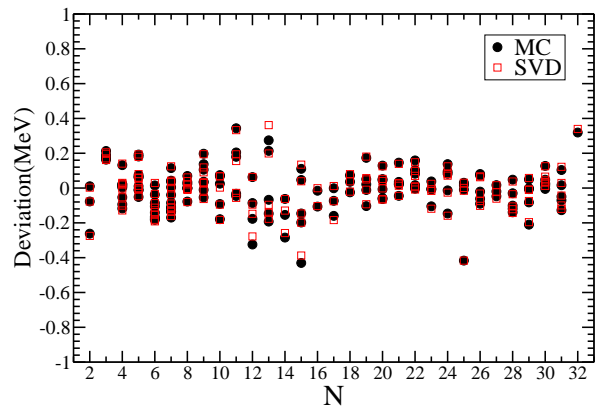}
\caption{(Color Online) Differences between experimental and calculated binding energies, $E_i^{\rm exp}-E_i^{\rm cal}$, as a function of valence neutron number.\label{diff}}
\end{figure}

The monopole Hamiltonians optimized with the SVD and MC approaches are similar to each other. To illustrate this point, in Fig. \ref{diff} we plotted the deviation from experimental data for calculations with the two effective Hamiltonians. The difference between the calculations is practically negligible. The largest deviation from experiments is seen at $N=15$ where the ground state energy of $^{115}$Sn is under-estimated by around 420 keV. In the following only calculations done with the MC optimized effective Hamiltonian are presented for simplicity. 

\begin{figure}[htdp]
\includegraphics[scale=0.35]{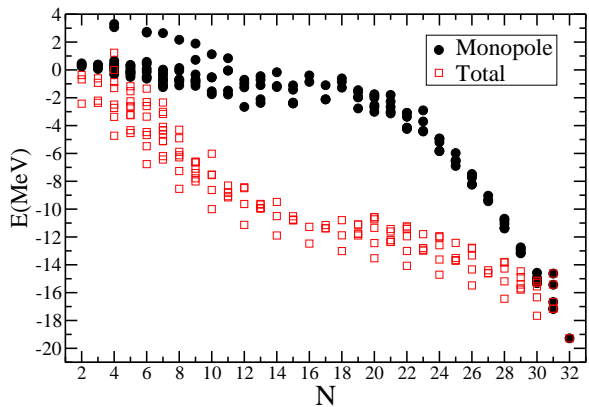}
\caption{(Color Online) The calculated shell model energies (solid), $E^{SM}=<H>$, and contributions from monopole Hamiltonian $<H_m>$ (open).\label{mono}}
\end{figure}

The calculated shell model energies, $E^{SM}=<H>$, of the selected states in tin isotopes are plotted in Fig. \ref{mono}. The contributions from the monopole Hamiltonian are also presented for comparison. From the figure one can see that the contribution from the multipole Hamiltonian reaches its maximum around the mid-shell.

\begin{figure*}[htdp]
\includegraphics[scale=0.54]{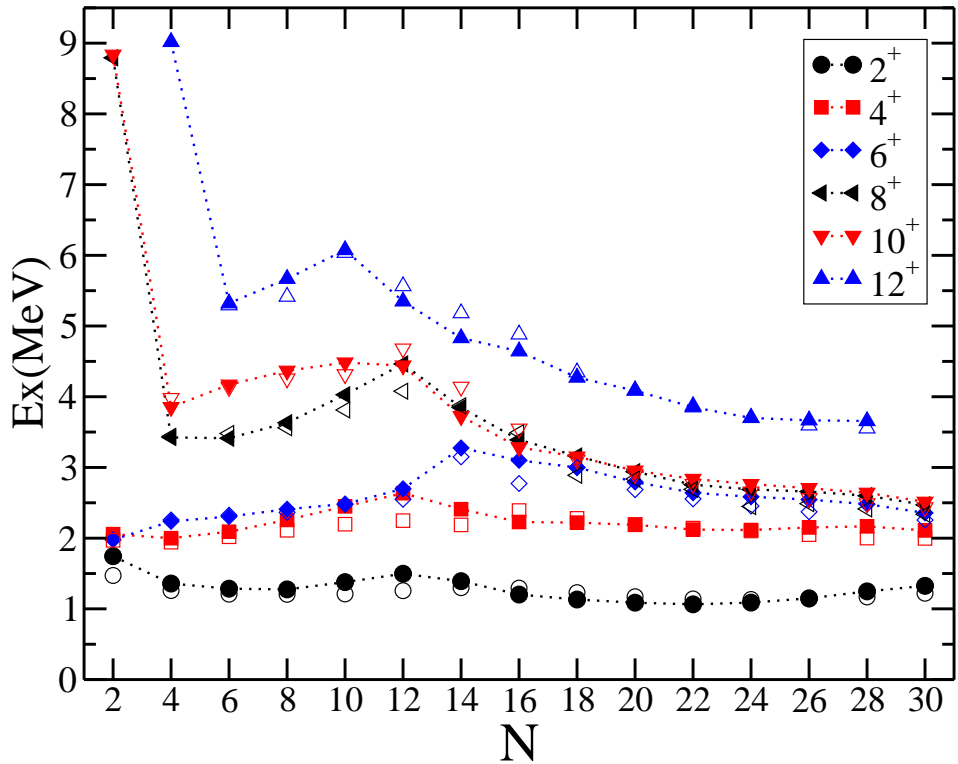}
\caption{(Color Online) Experimental \cite{nudat,Ast12,Pie11,Fot11} (open symbols) and calculated (solid symbols with dotted lines) excitation energies of the low-lying even-spin states in nuclei $^{102-130}$Sn.\label{even}}
\end{figure*}

The excitation energies of the 126 excited states can be reproduced within an average deviation of about 150 keV. 
The largest difference is seen in the $3/2^+_1$ state in $^{115}$Sn mentioned above, where the experimental datum is under-estimated by about 540 keV. Extensive experimental efforts are made recently to explore higher-seniority states built on the $10^+$ states in Sn isotopes \cite{Ast12,Pie11,Fot11}.
In Fig. \ref{even} we plot the calculated excitation energies of the low-lying even-spin states up to $I^{\pi}=12^+$ in the nuclei $^{102-130}$Sn. The experimental data are also plotted for comparison. They can be found in Refs. \cite{nudat,Ast12,Pie11,Fot11}.

\begin{figure}[htdp]
\includegraphics[scale=0.35]{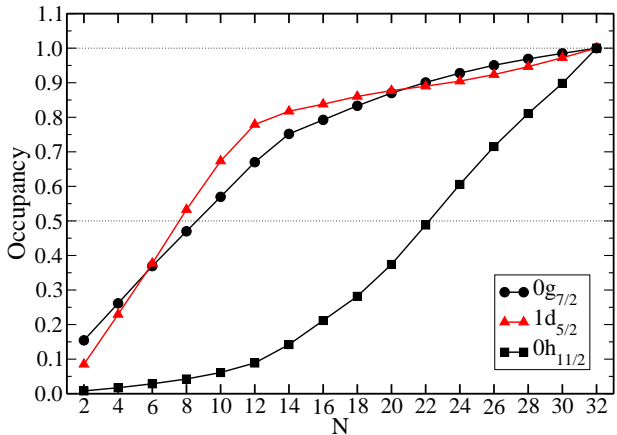}
\caption{(Color Online) Shell model occupancies, $<\hat{N}_j>/(2j+1)$, of the three higher-$j$ shells in the ground states of even tin isotopes.\label{occ}}
\end{figure}

\begin{table}[htdp]
  \centering
  \caption{Shell model occupancies, $<\hat{N}_j>/(2j+1)$, for the ground states of even tin isotopes $^{112-124}$Sn calculated in this work in comparison with experimental data \cite{ea,eb} and those calculated in Ref. \cite{And96}.}\label{table4}
\begin{ruledtabular}
  \begin{tabular}{ccccc}
$j$&Ref. \cite{ea}&Ref. \cite{eb}&Ref. \cite{And96}&This work\\
\hline
$A$&112\\
\hline
$5/2$&0.70&$0.93(12)$&0.76&0.78\\
$7/2$&0.69&0.63(13)&0.62&0.67\\
$1/2$&0.11&0.24(3)&0.30&0.21\\
$3/2$&0.14&0.18(3)&0.16&0.12\\
$11/2$&0.12&-&0.10&0.089\\
\hline
$A$&114\\
\hline
$5/2$&0.60&$0.97(9)$&0.87&0.82\\
$7/2$&0.86&0.69(15)&0.81&0.75\\
$1/2$&0.26&0.34(3)&0.28&0.29\\
$3/2$&0.31&0.37(4)&0.13&0.20\\
$11/2$&0.17&0.25(7)&0.11&0.14\\
\hline
$A$&116\\
\hline
$5/2$&0.81&$1.05(10)$&0.90&0.84\\
$7/2$&0.88&0.75(19)&0.86&0.79\\
$1/2$&0.52&0.60(5)&0.49&0.42\\
$3/2$&0.32&0.40(5)&0.20&0.31\\
$11/2$&0.15&0.30(7)&0.16&0.21\\
\hline
$A$&118\\
\hline
$5/2$&0.82&$1.0(1)$&0.88&0.86\\
$7/2$&0.81&0.78(19)&0.85&0.83\\
$1/2$&0.64&0.80(10)&0.48&0.55\\
$3/2$&0.38&0.60(8)&0.33&0.42\\
$11/2$&0.30&0.35(8)&0.30&0.28\\
\hline
$A$&120\\
\hline
$5/2$&0.94&$1.00(9)$&0.90&0.88\\
$7/2$&0.70&0.67(16)&0.88&0.87\\
$1/2$&0.70&0.95(10)&0.58&0.62\\
$3/2$&0.50&0.65(3)&0.43&0.51\\
$11/2$&0.42&0.38(9)&0.39&0.37\\
\hline
$A$&122\\
\hline
$5/2$&0.86&$1.00(9)$&0.92&0.89\\
$7/2$&-&0.58(16)&0.91&0.90\\
$1/2$&0.73&0.95(10)&0.67&0.66\\
$3/2$&0.51&0.65(9)&0.52&0.57\\
$11/2$&-&0.38(12)&0.48&0.49\\
\hline
$A$&124\\
\hline
$5/2$&0.94&$1.00(9)$&0.94&0.90\\
$7/2$&-&0.81(23)&0.93&0.93\\
$1/2$&0.80&0.95(10)&0.75&0.69\\
$3/2$&0.69&0.75(10)&0.62&0.62\\
$11/2$&-&0.38(12)&0.58&0.61\\
  \end{tabular}
  \end{ruledtabular}
\end{table}

In Fig. \ref{occ} we present the calculated occupancies of the single-particle orbitals in the ground state wave functions of even tin isotopes. The comparison with available experimental data \cite{ea,eb} (taken from Table I \& II in Ref. \cite{And96}) and calculations from Ref. \cite{And96} is presented in Table \ref{table4}. The structure of the low-lying states in light tin isotopes are dominated by configuration mixing between the orbitals $0g_{7/2}$ and $1d_{5/2}$. The two orbitals are half-filled around $N=8$ ($^{108}$Sn).
The orbital $0h_{11/2}$ is calculated to be half-filled around $N=22$ and 24. This is in agreement with the speculation in Ref. \cite{Bro92} by considering the E2 decay properties of the $10^+$ isomers in these nuclei.

\begin{figure}[htdp]
\includegraphics[scale=0.5]{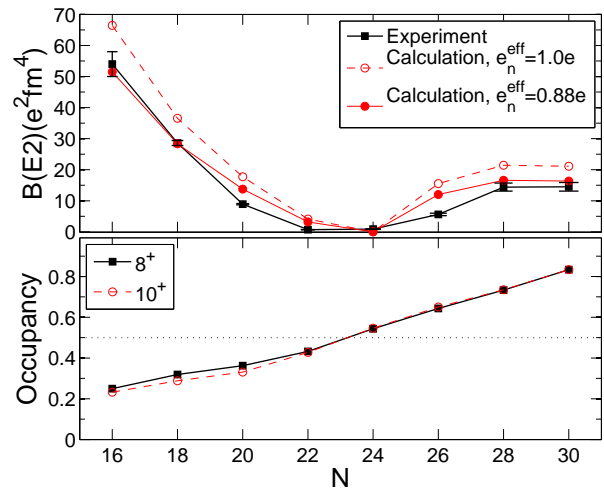}
\vspace{-2cm}
\caption{(Color Online) Experimental \cite{Bro92,Zhang00} and calculated $B$(E2) values on the transitions of the $10^+_1$ states in even Sn isotopes. The lower panel gives the calculated occpancies of the $0h_{11/2}$ orbital in the $10^+$ and $8^+$ states. \label{plus}}
\end{figure}

It is noticed in Ref. \cite{Bro92} that the $B$(E2) values practically vanish for the transitions between the $10^+_1$ and $8^+_1$ states in $^{122,124}$Sn. In that paper this is analyzed within the framework of the BCS approximation. That is, the $B$(E2) value will minimize when the $0h_{11/2}$ orbital is half-filled. Such a scheme is indeed supported by our shell model calculations, as seen in Fig. \ref{plus}. In that figure the $B$(E2) values are calculated with two sets of neutron effective charge, $e^{{\rm eff}}_{n}=1.0e$ \cite{Banu05} and $e^{{\rm eff}}_{n}=0.88e$ \cite{Bro92}. As in Ref. \cite{Bro92}, a better agreement is obtained with $e^{{\rm eff}}_{n}=0.88e$.

The calculated excitation energies of the $11/2^-$, $1/2^+$, $5/2^+$ and $7/2^+$ one-neutron-hole states in $^{131}$Sn, relative to the $3/2^+$ state, are -0.022, 0.475, 1.720 and 2.521 MeV, respectively. These are in fair agreement with the experimental data \cite{Fog04,nudat}.

In Table \ref{table5} we present the comparison between experimental and calculated excitation energies and magnetic dipole moments of the low-lying states in odd-$A$ tin isotopes. The experimental data on excitation energies are taken from Ref. \cite{nudat,Fog04} while the magnetic moments are taken from the compilation in Ref. \cite{Sto05}.
Two kinds of calculations are presented. The first one (labeled by I) corresponds to calculations with the free $g$ factors of neutron, $g_s=-3.826$ and $g_l=0$. The last column corresponds to calculations with the effective $g$ factor, $g_s^{{\rm eff}}=0.7g_s$, where $q_s=0.7$ stands for an effective qunching factor. As can be seen from the Table, a much better agreement with experiment is obtained with the introduction of the quenching factor. The same quenching factor $q_s=0.7$ was also used  in Ref. \cite{hon09} for calculations in the $fpg$ shell and in Ref. \cite{Wal11} for the calculations of $^{127,128}$Sn.

\begin{table}[htdp]
  \centering
  \caption{Experimental \cite{Sto05} and calculated excitation energies (in MeV) and magnetic moments (in $\mu_N$) of the low-lying states in odd-$A$ Sn isotopes. Columns denoted by I and II correspond to calculations with the free and effective $g$ factors, respectively.}\label{table5}
\begin{ruledtabular}
  \begin{tabular}{ccccccc}
Nucl.&$I^{\pi}$&$E_x^{{\rm Expt.}}$&$E_x^{{\rm Cal.}}$&$\mu^{{\rm Expt.}}$&$\mu^{{\rm Cal.}}$(I)&$\mu^{{\rm Cal.}}$(II)\\
\hline
$^{103}$Sn&$5/2^+$&0&0&-&-1.856&-1.299\\
&$7/2^+$&0.168&0.155&-&1.449&1.014\\
$^{105}$Sn&$5/2^+$&0&0&-& -1.723&-1.206\\
&$7/2^+$&0.1997&0.202&-&1.418&0.993\\
$^{107}$Sn&$5/2^+$&0&0&-& -1.591&-1.114\\
&$7/2^+$&0.151&0.241&-&1.345&0.942\\
$^{109}$Sn&$5/2^+$&0&0&-1.079(6)& -1.463&-1.024\\
&$7/2^+$&0.0135&0.221&-&1.320&0.924\\
$^{111}$Sn&$7/2^+$&0&0&0.608(4)& 1.246&0.872\\
&&&&0.617(8)\\
&$5/2^+$&0.154&-0.083&-&-1.353&-0.947\\
&$11/2^-$&0.979&0.766&-1.26(11)&-1.741&-1.219\\

$^{113}$Sn&$1/2^+$&0&0&-0.8791(6)&-1.009&-0.706\\
&$3/2^+$&0.498&0.0311&-&0.769&0.538\\
&$5/2^+$&0.4098&0.0271&-&-0.455&-0.318\\
&$7/2^+$&0.077&0.0314&-&1.278&0.895\\
&$11/2^-$&0.738&0.332 &-1.30(2)&-1.710&-1.197\\
&&&&-1.29(2)\\

$^{115}$Sn&$1/2^+$&0&0&-0.91883(7)&-1.088&-0.762\\
&$3/2^+$&0.497&-0.045&-&0.793&0.555\\
&$5/2^+$&0.987&0.724&-&-1.317&-0.922\\
&$7/2^+$&0.613&0.403&0.683(10)&1.366&0.914\\
&$11/2^-$&0.714&0.237 &-1.378(11)&-1.745&-1.222\\
&&&&-1.369(4)\\

$^{117}$Sn&$1/2^+$&0&0&-1.00104(7)&-1.238&-0.867\\
&$3/2^+$&0.159&-0.058&0.66(5)&0.790&0.553\\
&$7/2^+$&0.7115&0.572&-&1.321&0.925\\
&$11/2^-$&0.315&0.171 &-1.3955(10)&-1.791&-1.253\\

$^{119}$Sn&$1/2^+$&0&0&-1.04728(7)&-1.268&-0.888\\
&$3/2^+$&0.0239&-0.131&0.633(3)&0.854&0.598\\
&&&&0.682(3)\\
&$11/2^-$&0.0895&0.128 &-1.40(8)&-1.825&-1.278\\
$^{121}$Sn&$3/2^+$&0&0&0.6978(10)& 0.923&0.646\\
&$11/2^-$&0.0063&-0.0086&-1.3877(9)&-1.844&-1.291\\
$^{123}$Sn&$11/2^-$&0&0&-1.3700(9)&-1.854&-1.298\\
&$3/2^+$&0.025&0.0703&-&1.005&0.703\\
$^{125}$Sn&$11/2^-$&0&0&-1.348(6)&-1.866&-1.306\\
&$3/2^+$&0.028&0.053&0.764(3)&1.071&0.750\\
$^{127}$Sn&$11/2^-$&0&0&-1.329(7)&-1.885&-1.319\\
&$3/2^+$&0.0047&-0.027&0.757(4)&1.106&0.774\\
$^{129}$Sn&$3/2^+$&0&0&0.754(6)&1.128&0.789\\
&$11/2^-$&0.035&0.0912&-1.297(5)&-1.907&-1.335\\
$^{131}$Sn&$3/2^+$&0&0&0.747(4)&1.148&0.804\\
&$11/2^-$&0.069(14)&-0.0225&-1.276(5)&-1.913&-1.339
\\
  \end{tabular}
  \end{ruledtabular}
\end{table}

\begin{figure}[htdp]
\includegraphics[scale=0.45]{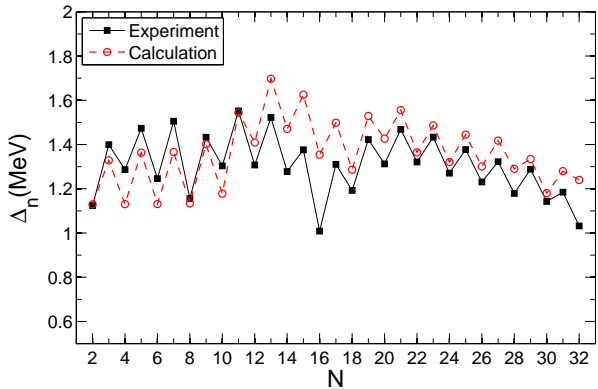}
\caption{(Color Online) Neutron pairing gaps in Sn isotopes extracted from the experimental and calculated binding energies. \label{delta}}
\end{figure}

The empirical pairing gaps can readily
be obtained from the experimental and calculated binding energies as
\begin{equation}\label{dexp}
\Delta_n(N)=\frac{(-1)^{(N+1)}}{2}\left[{\rm BE}(N)+{\rm BE}(N-2)-2{\rm BE}(N-1)\right].
\end{equation}
These gaps are shown as a function of the neutron number in Fig. \ref{delta}. As can be seen from the figure, the overall agreement between experiments and calculations on the pairing gaps are quite satisfactory. Noticeable differences are only seen for mid-shell nuclei $^{114-117}$Sn. This is related to the relatively large difference between experimental and calculated binding energies of $^{115}$Sn. It may indicate that the $J=0$ pairing matrix elements  in the CD-Bonn potential involving the $2s_{1/2}$ and $1d_{3/2}$ orbitals may not be perfectly described.

\section{Spin inversion in $^{103}${Sn}}

The spins of the ground state and first excited state in $^{103}${Sn} are $I=5/2$ and 7/2, respectively \cite{Fah01}, which are reversed with respect to those in  $^{101}${Sn} \cite{Dar10}. 
Through seniority model analyses with a pairing Hamiltonian, Ref.~\cite{Dar10} suggested that the inversion is dominated by orbital-dependent pairing correlations, namely the strength of the pairing matrix elements $\langle0g_{7/2}^2|V|0g_{7/2}^2\rangle_{J=0}$ is much larger than that of $\langle1d_{5/2}^2|V|1d_{5/2}^2\rangle_{J=0}$. This produces strong additional binding for the $J^{\pi}=5/2^+$ state in $^{103}${Sn}, which eventually becomes the ground state. The effect of other interaction terms on the spin inversion was not considered in Ref. \cite{Dar10}.

\begin{figure}[htdp]
\includegraphics[scale=0.35]{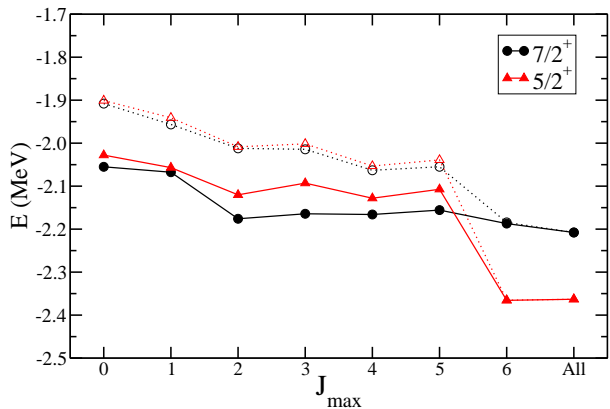}
\caption{(Color Online) The calculated shell-model energies of the first $5/2^+$ and $7/2^+$ states in $^{103}$Sn with a Hamiltonian $H'$ containing the single-particle terms and two-body matrix elements with $J\leq J_{\rm max}$. The solid symbols denote the eigenvalues of $H'$ derived by diagonalizing the Hamiltonian matrix. The open symbols denote the expectation values of $H'$, $\langle\Psi_I|H'|\Psi_I\rangle$, with respect to the corresponding wave functions $|\Psi_I\rangle$ of the full Hamiltonian $H$,  where all matrix elements are taken into account.\label{sn103}}
\end{figure}

In Fig. \ref{sn103} we analyze the contribution from different components of the effective Hamiltonian to the spin inversion between $^{101}$Sn and $^{103}$Sn. For that purpose the energies of the first $J^{\pi}=5/2^{+}$ and $7/2^{+}$ states are calculated with a limited Hamiltonian $H'$ containing the single-particle terms and two-body matrix elements with $J\leq J_{\rm max}$ only.  $J_{\rm max}$ denotes the maximal spin value of the two-body matrix elements to be considered. 
Two different types of calculations are presented in the figure.
The solid symbols correspond to the results calculated by diagonalizing the Hamiltonian $H'$.
The expectation values of such a Hamiltonian with respect to the corresponding wave functions $|\Psi_I\rangle$ of the full Hamiltonian $H$, $\langle\Psi_I|H'|\Psi_I\rangle$, are plotted as open symbols. It should be mentioned that calculations with the original CD-Bonn interaction and other realistic nucleon-nucleon potentials give a similar pattern.

It is thus seen from Fig. \ref{sn103} that both calculations give similar results concerning the order of the $5/2^{+}$ and $7/2^{+}$ states. Calculations with the pairing matrix elements only (i.e.,  $J_{\rm max}=0$) show that the pairing terms, in particular the element $\langle0g_{7/2}^2|V|0g_{7/2}^2\rangle_{J=0}$ mentioned above,  can significantly reduce the gap between the two states but were not strong enough to induce the inversion. A sudden switch is seen when the $J=6$ two-body matrix elements are considered. 
It can be seen from Fig. \ref{sn103} that in both calculations the exact results are also approached by including terms with $J\leq6$ only.
This is expected since the low-lying states of light tin isotopes mainly occupy the nearly degenerate orbitals $0g_{7/2}$ and $1d_{5/2}$ for which the maximal spin is $J=6$. Thus the contribution from interactions with higher spin values are marginal. Among $T=1$ matrix elements the maximal spin one can have is $J=10$. It corresponds to the coupling of two nucleons in the orbital $0h_{11/2}$. Calculations in the restricted $0g_{7/2}1d_{5/2}$ model space give a result similar to Fig. \ref{sn103}. 

Among the $J=6$ two-body matrix elements the ones that contribute most to the spin inversion are the repulsive matrix element $\langle0g_{7/2}^2|V|0g_{7/2}^2\rangle_{J=6}$ and the strongly attractive one $\langle0g_{7/2}1d_{5/2}|V|0g_{7/2}1d_{5/2}\rangle_{J=6}$. This can be understood by considering the structure of the wave functions of the two states. In the $5/2^+$ ground state in $^{103}$Sn, the leading component is $|(0g_{7/2}^2)_{J=0}1d_{5/2}\rangle_I$ \cite{Dar10}. Its overlap with the total wave function is calculated to be
$|\langle(0g_{7/2}^2)_{J=0}1d_{5/2}|\Psi\rangle_I|=0.86$. One can also construct a three-body state starting from the pair $|0g_{7/2}1d_{5/2}\rangle_{J=6}$. The overlap between the state thus constructed and the total wave function is $|(0g_{7/2}1d_{5/2})_{J=6}0g_{7/2}|\Psi\rangle_I=0.75$.
The $\langle0g_{7/2}1d_{5/2}|V|0g_{7/2}1d_{5/2}\rangle_{J=6}$ term induces a significant additional binding  for the $J^{\pi}=5/2^+$ state in $^{103}${Sn}, as can be seen from Fig. \ref{sn103}.
It should be mentioned that states generated by the two couplings $|(0g_{7/2}^2)_{J=0}1d_{5/2}\rangle$ and $(0g_{7/2}1d_{5/2})_{J=6}0g_{7/2}\rangle_I$ are not perpendicular to each other. Their overlap is quite large, $\langle(0g_{7/2}^2)_{J=0}1d_{5/2}|(0g_{7/2}1d_{5/2})_{J=6}0g_{7/2}\rangle_I=0.74$. This can be evaluated analytically \cite{Qi10}.

The overlaps of the total wave function of the first $7/2^+$ state $^{103}$Sn with its leading components are calculated to be $|\langle(1d_{5/2}^2)_{J=0}0g_{7/2}|\Psi\rangle_I|=0.65$, $|\langle(0g_{7/2}^2)_{J}0g_{7/2}|\Psi\rangle_I|=0.62$ and $|\langle(0g_{7/2}1d_{5/2})_{J=6}1d_{5/2}|\Psi\rangle_I|$=0.57.
From a shell-model point of view, the couplings $|(0g_{7/2}^2)_{J=0} 0g_{7/2}\rangle_I$ and  $|(0g_{7/2}^2)_{J=6} 0g_{7/2}\rangle_I$ generate exactly the same three-particle state. All interaction terms $\langle0g_{7/2}^2|V|0g_{7/2}^2\rangle_{J}$ contribute to the total energy of the state \cite{Qi10}. It may be interesting to mention that the effect of the maximally aligned pair in single-$j$ systems was discussed in Refs. \cite{Chen92,Qi11}.

\begin{figure}[htdp]
\vspace{0.5cm}
\includegraphics[scale=0.35]{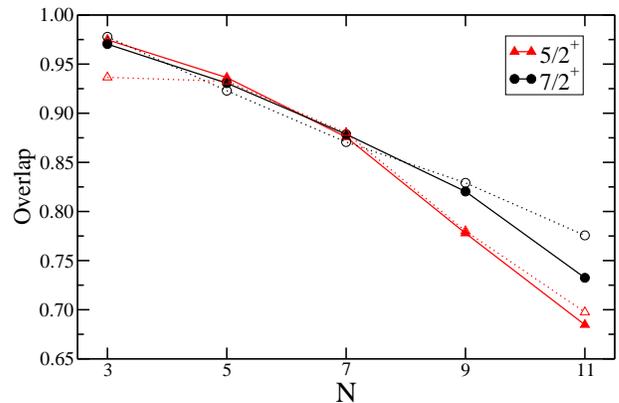}
\caption{(Color Online) Solid symbol: The overlaps between the wave functions $|\Psi_I\rangle$ of the full Hamiltonian $H$ and those of the pairing Hamiltonian with $J_{\rm max}=0$ for the first $5/2^+$ and $7/2^+$ states in light odd-$A$ Sn isotopes. Open symbol: Same as above but only the non-diagonal pairing matrix elements are considered.\label{odd}}
\end{figure}

The influence of the pairing interaction on the structure of tin isotopes was considered in a variety of approaches (see, e.g., \cite{And96,Dean03}).
In Fig. \ref{odd} we calculated the overlaps between the full shell-model wave functions and those calculated from the pairing Hamiltonian including the single-particle terms and the $J=0$ pairing matrix elements. It is thus seen that the non-diagonal pairing matrix elements $\langle j^2|V|j'^2\rangle_{J=0}$ play an essential role in inducing the configuration mixing in the first $5/2^+$ and $7/2^+$ states of odd-$A$ tin isotopes. The overlap gradually decreases when the number of valence neutrons increases. This is consistent with the generalized seniority model calculations in Ref. \cite{San97}, namely the overlap between the full wave functions and seniority-truncated state also decreases with increasing neutron number.  

\section{Summary}
The structure properties of light tin isotopes are calculated with a global optimized effective interaction. The unknown single-particle energies of the orbitals $1d_{3/2}$, $2s_{1/2}$ and $0h_{11/2}$ and the monopole interactions are refined by fitting to experimental binding energies. A total number of 157 states in $^{102-132}$Sn are considered in the fitting. The binding energies of these states can be reproduced within an average deviation of about 130 keV. The largest deviation is around 600 keV which is seen in the nucleus $^{115}$Sn. With the effective Hamiltonian thus derived we calculate the contributions of the monopole and multipole Hamiltonian to the total shell-model energies. The excitation energies of the low-lying even spin states in even Sn isotopes are presented. We also evaluated the shell model occupancies in the ground states of these nuclei.
Detailed systematic calculations on the spectra and decay properties of tin isotopes as well as the list of the two-body matrix elements will be presented in Ref. \cite{sn100}.

We analyze the origin of the spin inversion between the ${7/2}^+$ and $5/2^+$ states in $^{103}$Sn and heavier odd tin isotopes in order to explore the possible influence of different interaction channels. We thus find that both the $J=0$ pairing and  the maximally aligned $J=6$ two-body matrix elements produce strong additional binding for the $5/2^+$ states. The non-diagonal pairing matrix elements play an essential role in inducing the mixing of different  configurations in the wave functions of these states.

Within the framework as described in this paper, we have done a preliminary optimization of the $T=0$ monopole interaction by fitting to the binding energies of Sb, Te and I isotopes around the $N=50$ and 82 shell closures. This will be available in Ref. \cite{sn100}.

\section*{Acknowledgment}
We thank T. B\"ack, R. Liotta and R. Wyss for stimulating discussions.
This work has been supported by the Swedish Research Council (VR) under grant No. 621-2010-4723. CQ also acknowledges the computational support from the Swedish National Infrastructure for Computing (SNIC) at
PDC and NSC.

\end{document}